\documentclass{Interspeech2024}
\usepackage{pifont}
\usepackage{multirow}

\interspeechcameraready

\title{Improving Whisper's Recognition Performance for Under-Represented Language Kazakh Leveraging Unpaired Speech and Text}

\name{Jinpeng}{Li}
\name{Yu}{Pu}
\name{Qi}{Sun}
\name[affiliation={*}]{Wei-Qiang}{Zhang}

\address{
Department of Electronic Engineering, Tsinghua University, Beijing 100084, China}
\email{\thanks{* Corresponding author}\thanks{This work was supported by the National Natural Science Foundation of China under Grant No. 62276153.}lijp22@mails.tsinghua.edu.cn,wqzhang@tsinghua.edu.cn}


\newcommand{\red}[1]{\textcolor{red}{#1}}

\begin{document}
\maketitle

\begin{abstract}
    
Whisper and other large-scale automatic speech recognition models have made significant progress in performance. However, their performance on many low-resource languages, such as Kazakh, is not satisfactory. It is worth researching how to utilize low-cost data to improve the performance of Whisper on under-represented languages. In this study, we utilized easily accessible unpaired speech and text data and combined the language model GPT with Whisper on Kazakh. We implemented end of transcript (EOT) judgment modification and hallucination penalty to improve the performance of speech recognition. Further, we employed the decoding average token log probability as a criterion to select samples from unlabeled speech data and used pseudo-labeled data to fine-tune the model to further improve its performance. Ultimately, we achieved more than 10\% absolute WER reduction in multiple experiments, and the whole process has the potential to be generalized to other under-represented languages.

\end{abstract}

\section{Introduction}

The development of end-to-end (E2E) automatic speech recognition (ASR) systems has seen significant advancements in the field of speech recognition \cite{li2022recent, radford2023robust, zhang2023google,ZhaoZY2021_NN}. However, training E2E models to achieve satisfactory recognition results requires large amounts of high-quality labeled speech data \cite{kunze2017transfer}. This poses a substantial bottleneck for low-resource languages that lack adequate labeled speech data. Whisper \cite{radford2023robust} is a universal multilingual speech recognition model trained on 680,000 hours of supervised data. However, it has not been as effective for many low-resource languages. Since only a few languages have sufficient annotated speech data, while most languages are resource-scarce in this regard, it is crucial to investigate how limited low-cost data can be leveraged to improve low-resource ASR systems.

Various strategies have been proposed by researchers to address this challenge. Multilingual transfer learning and multilingual meta-learning \cite{li2021scaling, dalmia2018sequence, 
qian2022optimizing} are two approaches that utilize labeled data to pre-train a foundational model that can be applied across multiple languages. However, both methods require paired labeled data in both the source and target languages for pre-training and fine-tuning, respectively. Unfortunately, paired data for under-represented languages is scarce.

A promising alternative lies in leveraging unlabeled data through self-supervised or semi-supervised learning techniques. Self-supervised learning (SSL) leverages readily available unpaired speech data. Inspired by masked language models in text, masked acoustic models are trained to predict masked segments of speech, learning representations without labels \cite{devlin2018bert, liu2020mockingjay, liu2021tera}. These SSL models, when fine-tuned with a small amount of labeled data, have shown significant improvements in low-resource ASR systems. The success of models like wav2vec2 XLSR-53 and HuBERT, pre-trained on vast amounts of unlabeled data, exemplifies the effectiveness of SSL in this domain \cite{conneau2020unsupervised,hsu2021hubert,ZhaoJ2022_JSTSP}. However, self-supervised methods are difficult to apply to the trained Whisper model due to the fact that its encoder already has excellent representation capabilities through large-scale supervised training. In this case, semi-supervised methods are more suitable, such as iterative pseudo-labeling, which utilizes language models to create pseudo-labels for unlabeled data and combines them with a small amount of labeled data to expand the training set \cite{xu2020iterative,park2020improved}. 

In this paper, we select Kazakh as an example of under-represented languages for study.
Kazakh is the official language of Kazakhstan and belongs to the Turkic language family, but it is still under-represented in speech recognition. Whisper does not perform well in Kazakh, specifically, the word error rate (WER) for Kazakh is over 40\% on the Fleurs \cite{conneau2023fleurs} test set and over 55\% on the KSC \cite{mussakhojayeva2022ksc2} test set. In this work, we only utilize low-cost unpaired speech and text data that can be easily accessed online without requiring manual labeling. This appears to be related to approximate unsupervised learning methods, such as wav2vec-U \cite{baevski2021unsupervised,liu2023towards}. However, due to the requirement of pronunciation lexicon information and the difficulty in achieving convergence during training of GAN models, we chose to use this data to improve the performance of Whisper.



Specifically, we leveraged easily accessible text data and integrated language model GPT with Whisper, implementing improvements such as end of transcript (EOT) judgment modification and hallucination penalty. We found that decoding with GPT leads to a more significant reduction in WER for samples with higher average token log probability (ALP). Therefore, we employed ALP as a criterion to select samples from unlabeled speech data, thus fine-tuning the model using pseudo-labels. In multiple experiments, we achieved more than 10\% absolute WER reduction, and the pipeline is scalable to other under-represented languages.

\section{Methods}
\subsection{Leveraging text data}
\subsubsection{Integrate GPT with Whisper}

We first trained a language model using text data and then integrated it as an additional decoder in the Whisper framework, working in conjunction with its original decoder. Through this approach, our goal is to harness the complementary advantages of the two decoders to improve Whisper's recognition performance on under-represented languages. Similar attempts have been made previously by integrating the official GPT-2 model into the decoding process of Whisper and applying it to English \cite{sun2023can}. However, since Whisper already performs well in English, the improvement was relatively slight. In our study, we utilized mGPT \cite{shliazhko2024mgpt} with 1.3 billion parameters, sharing the same architecture as GPT-3 \cite{brown2020language}. After adjusting the tokenizer to be consistent with Whisper and retraining it with text data, we applied this model to under-represented languages to improve recognition performance.

\begin{figure}[htp]
  \centering
  \includegraphics[width=0.99\linewidth]{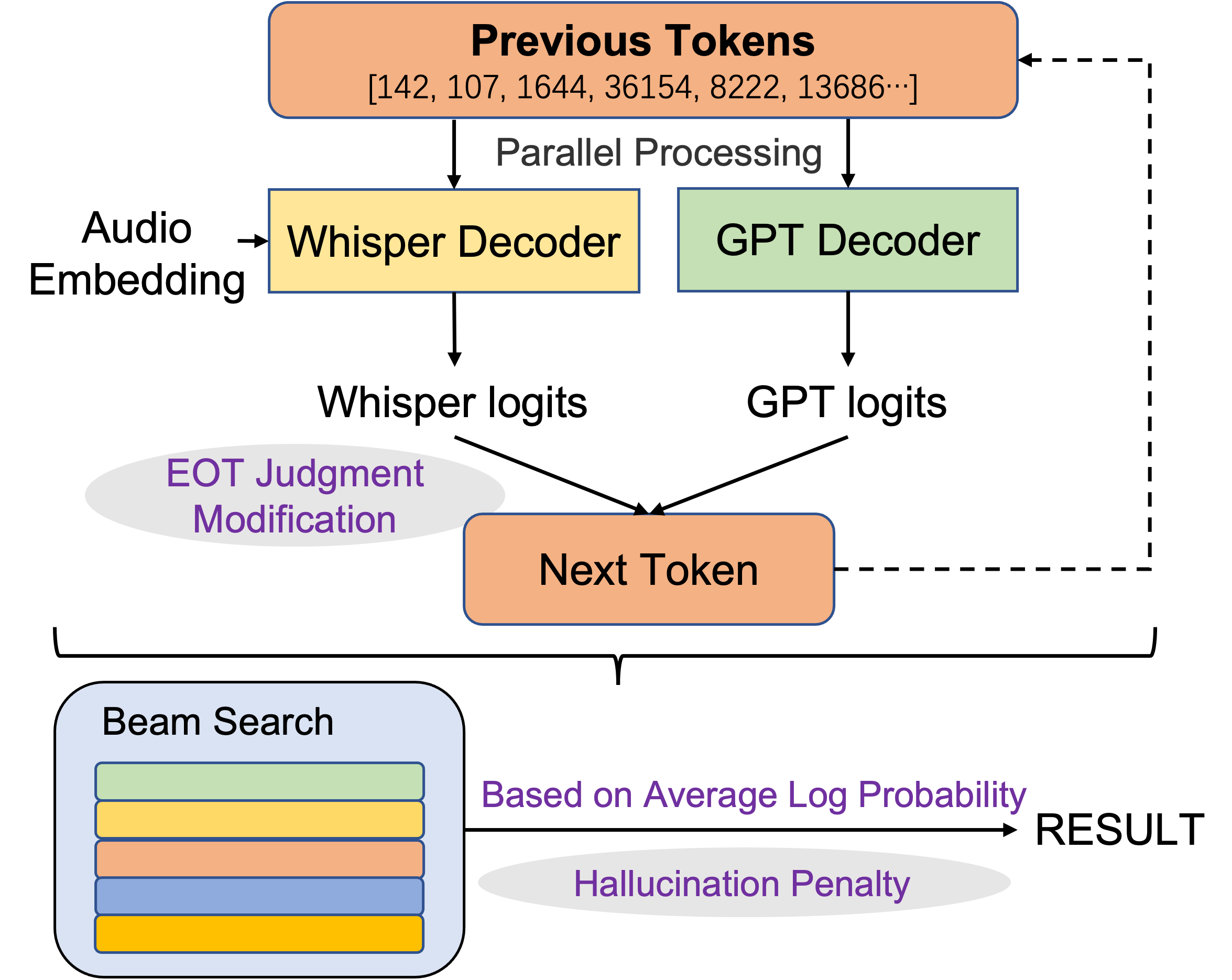}
  \caption{Integrating GPT into the decoding process of Whisper.}
  \label{fig:gpt}
\end{figure}

Figure \ref{fig:gpt} illustrates the decoding process after integrating GPT into Whisper. Let the audio input be denoted as $X$, and in a certain iteration of the autoregressive decoding, the token sequence of the preceding context is denoted as $T$. The probabilities for the next decoding token $Y$ for Whisper decoder and GPT are $P_{\text{Whisper}}(Y|X,T)$ and $P_{\text{GPT}}(Y|T)$, respectively. The weight of GPT is defined as $\lambda_{\text{GPT}}$. The selection criterion for $Y$, i.e., the calculation formula for the log probability (LP) of this token, is given by:
\begin{equation}
    \mathrm{LP} = \frac{1}{1+\lambda_{\text{GPT}}}(\log P_{\text{Whisper}}(Y|X,T) + \lambda_{\text{GPT}} \log P_{\text{GPT}}(Y|T))
\end{equation}
In the equation, we found that taking the logarithm of the two probabilities separately and then adding them weightedly yields better results than taking the weighted sum of probabilities and then taking the logarithm. In addition, when determining the end of transcription, it is important to depend on the speech content and not allow the language model to generate redundant content. Therefore, we made end of transcription (EOT) judgment modification: based on the output probability of Whisper, if the current token's probability of being EOT is the highest, then $\lambda_{\text{GPT}}$ is set to 0. Experiments have confirmed the importance of this improvement.

\subsubsection{Hallucination penalty}
During the decoding process, instances where the interruption occurs due to the number of tokens surpassing the predetermined upper limit often indicate potential issues such as hallucinations or excessively lengthy audio segments. Such occurrences pose a significant risk of diminishing the credibility of the resultant transcript. To address this concern, we introduce a penalty mechanism aimed at mitigating the adverse effects of excessively long transcripts. Specifically, when this situation occurs, we impose the following penalty on the sum of token log probabilities (SLP):
\begin{equation}
  \mathrm{SLP}:= \mathrm{SLP} - N \cdot \log(2)
\end{equation}
where $N$ denotes the number of decoding tokens. SLP is typically used to divide the number of decoding tokens to get the average log probability (ALP) as the criterion for final selection in beam search. The penalty is intended to approximate the halving of the probabilities of all tokens when decoding is interrupted due to the token limit being exceeded.

On the other hand, hallucinations are often presented in the form of sentence repetition. Therefore, for all beam search candidates, we identify token strings in the sequence that exhibit cyclic behavior. The maximum length of the cyclic substring is denoted as $L$, and the number of repetitions is denoted as $C$. For instance, in the sequence \emph{ABCDABCD}, $L$ is 4 and $C$ is 1. Based on this, we introduce an additional penalty mechanism for SLP:
\begin{equation}
  \mathrm{SLP}:= \mathrm{SLP} - L \cdot C \cdot \log(2)
\end{equation}
This penalty is aimed at reducing the probability of additional tokens falling into a cycle, thus biasing the final selection in beam search towards candidates without hallucinations.

\subsection{Leveraging speech data}
After completing the decoding process, we recorded the text of each sample and its average token log probability (ALP), and found a certain correlation between the individual sample's WER and ALP. Furthermore, by incorporating GPT into the decoding process, we have a greater opportunity to predict sample decoding quality through ALP, leveraging the rich linguistic information provided by GPT. Therefore, we utilized unlabeled target domain speech data, decoded it with GPT to generate pseudo-labels, and then recorded the ALP. Subsequently, we sorted the samples based on ALP, and selected a subset with higher ALP values for fine-tuning the Whisper model. Through this pseudo-label fine-tuning approach, the Whisper model can better adapt to the target domain, learn from the knowledge of the language model, and improve speech recognition capability in under-represented languages.

\section{Experimental Setup}

\subsection{Details of leveraging text data}

The text dataset consists of the Kazakh portion from the Leipzig\footnote{https://huggingface.co/datasets/kz-transformers/multidomain-kazakh-dataset/blob/main/leipzig.csv} and Uzbek-Kazakh parallel corpora\footnote{https://huggingface.co/datasets/Sanatbek/uzbek-kazakh-parallel-corpora}, comprising approximately 1.7 million entries, utilized to train a 1.3B GPT-3 model. The model is initialized with the parameters of mGPT-Kazakh \cite{shliazhko2024mgpt}, setting the tokenizer to be consistent with Whisper's multilingual model. The model is trained for one epoch using the text dataset, with the optimizer set to AdamW and the scheduler set to WarmupLR. During decoding, we set the beam search size to 5, consistent with the default value of the transcribe interface. In the decoding loop, we use key-value cache to record intermediate key-value pairs generated during the decoding process to accelerate the decoding speed.

To explore the effect of model size on the results, we employed two Whisper models of different scales: Whisper-base with 74 million parameters and Whisper-large with 1.5 billion parameters. As the former exhibited a WER over 100\% on Kazakh, we fine-tuned it using the Kazakh dataset from Fleurs \cite{conneau2023fleurs}, resulting in the model called Whisper-base-KF. This fine-tuned model achieved a WER on Kazakh similar to that of the original Whisper-large model. Whisper-base-KF and Whisper-large have GPT weights of 0.3 and 0.1 at decoding, respectively.

\subsection{Details of leveraging speech data}

For the unlabeled speech data, approximately 500 hours of the crowdsourced portion (KSC) from the KSC2 dataset \cite{mussakhojayeva2022ksc2} were employed to generate pseudo-labels for Whisper-base-KF, while around 10 hours of the Kazakh training set from Fleurs were utilized to generate pseudo-labels for Whisper-large. During the fine-tuning of Whisper with pseudo-labels, we froze the encoder and only fine-tuned the decoder. We used the cross-entropy loss function and employed the AdamW optimizer. The hyperparameters were set as follows: epoch: 5, batch size: 16, learning rate: 0.0001, and weight decay: 0.01.


\section{Results}

\subsection{Leveraging text data}

After training the GPT model using text data, we conducted tests to measure its perplexity (ppl) on the Fleurs and KSC test sets. The perplexity results for the Fleurs and KSC test sets were 2.61 and 6.20, respectively. Furthermore, the trained GPT model was integrated into Whisper, and the decoding results of the Whisper-base-KF and whisper-large models on the Fleurs and KSC test sets are shown in Table \ref{tab:gpt2}.

\begin{table}[h]
  \caption{Summary of Word Error Rate (WER) of the two models with/without GPT in Fleurs and KSC test sets.}
  \label{tab:gpt2}
  \centering
  \scalebox{0.95}{ 
  \begin{tabular}{lcc}
    \toprule
    \textbf{Model}  & \textbf{Fleurs WER(\%)} & \textbf{KSC WER(\%)}             \\
    \midrule
    Whisper-base-KF & 37.31 & 61.51 \\
    \quad+GPT for decoding & 28.60 & 50.53 \\
    \midrule
    Whisper-large & 43.58 & 56.18 \\
    \quad+GPT for decoding & 36.64 & 49.24 \\
    \bottomrule
  \end{tabular}
  }
\end{table}

It can be seen that integrating a well-trained GPT for the Kazakh language into the decoding process of Whisper, along with the utilization of EOT modification and hallucination penalty, can significantly improve the performance of both Whisper models. The gain brought by GPT for the speech recognition results is dependent on the scale of the Whisper model. For the relatively smaller Whisper model, Whisper-base-KF, the combination with the 1.3B GPT yields greater benefits, achieving a relative WER reduction of 23.3\% and 18.9\% on the Fleurs and KSC test sets, respectively. For the larger Whisper model, whisper-large, the combination with GPT results in a relative WER reduction of 15.9\% and 12.4\% on the Fleurs and KSC test sets, respectively.

\subsubsection{Impact of the modifications on the results}

\begin{table}[h]
  \caption{Results of models decoding with GPT, EOT Judgment Modification (EOT-JM), and Hallucination Penalty (HP) on Fleurs-test.}
  \label{tab:ablation}
  \centering
  \scalebox{0.95}{ 
  \begin{tabular}{lcccc}
    \toprule
    \textbf{Model} & \textbf{GPT} & \textbf{EOT-JM} & \textbf{HP} & \textbf{WER(\%)} \\
    \midrule
    \multirow{4}{*}{Whisper-base-KF} & \ding{55} & \ding{55} & \ding{55} & 37.31 \\
    ~ & \ding{51} & \ding{55} & \ding{55} & 34.49 \\
    ~ & \ding{51} & \ding{51} & \ding{55} & 28.78 \\
    ~ & \ding{51} & \ding{51} & \ding{51} & \textbf{28.60} \\
    \midrule
    \multirow{4}{*}{Whisper-large} & \ding{55} & \ding{55} & \ding{55} & 43.58 \\
    ~ & \ding{51} & \ding{55} & \ding{55} & 36.75 \\
    ~ & \ding{51} & \ding{51} & \ding{55} & 36.68 \\
    ~ & \ding{51} & \ding{51} & \ding{51} & \textbf{36.64} \\
    \bottomrule
  \end{tabular}
  }
\end{table}

Table \ref{tab:ablation} presents the Word Error Rate (WER) of two models with each improvement step on Fleurs-test. For the EOT Judgment Modification (EOT-JM) and Hallucination Penalty (HP), the smaller-scale Whisper-base-KF model with a higher language model weight exhibits a greater decrease in WER, particularly for the former improvement. This suggests that smaller-scale models rely more on larger language models during decoding, even for judging endings. EOT-JM ensures that the ending of transcription relies on audio information, reducing the generation of nonexistent information in the audio by GPT, resulting in a significant reduction in WER. The overall effect of HP on WER is not significant because HP affects only the candidate options and average token log probability (ALP) in the final beam search after decoding all tokens. However, for high-priority samples with higher ALP, HP can have a substantial impact.

\subsubsection{Impact of hallucination penalty on high-priority data}
\label{section:pdf_sanitise}

\begin{figure}[htp]
  \centering
  \includegraphics[width=\linewidth]{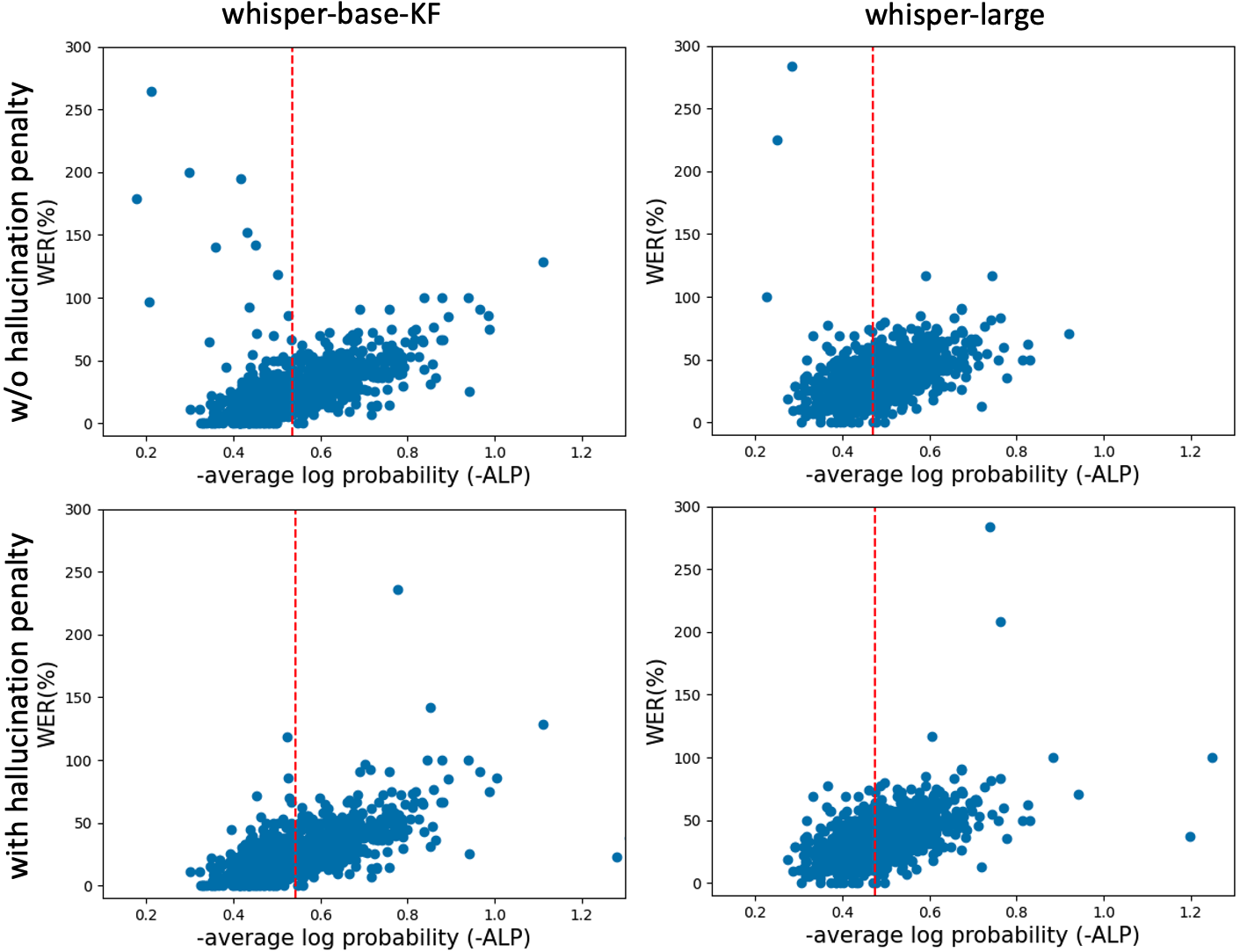}
  \caption{Decoded sample distribution of models on Fleurs-test. The X-axis represents the negative average log probability of the sample's tokens (-ALP), and the Y-axis represents the Word Error Rate (WER) for each sample. The red dashed line separates the samples into two halves based on -ALP.}
  \label{fig:sample}
\end{figure}

\begin{table}[h]
  \caption{Summary of WER for the Fleurs-test subset with a high average token log probability (ALP). The values highlighted in red represent the difference compared to the case without GPT.}
  \label{tab:topresult}
  \centering
  \scalebox{0.88}{ 
  \begin{tabular}{cccc}
    \toprule
    \textbf{Samples WER(\%)}  & \textbf{All} & \textbf{Top 20\% ALP} & \textbf{Top 50\% ALP}  \\
    \midrule
    \midrule
    \multicolumn{4}{l}{\textbf{Whisper-base-KF:}}\\
    \midrule
    w/o GPT & 37.31 & 22.17 & 29.32 \\
    \multirow{2}{*}{w/o HP} & 28.78 & 20.34 & 21.18 \\
    ~ & \textcolor{red}{(-8.53)} & \textcolor{red}{(-1.73)} & \textcolor{red}{(-8.14)} \\
    \multirow{2}{*}{with HP} & 28.60 & \textbf{12.42} & \textbf{18.07} \\
    ~ & \textcolor{red}{(-8.71)} & \textbf{\textcolor{red}{(-9.75)}} & \textbf{\textcolor{red}{(-11.25)}} \\
    \midrule
    \midrule
    \multicolumn{4}{l}{\textbf{Whisper-large:}}\\
    \midrule
    w/o GPT & 43.58 & 34.55 & 38.14 \\
    \multirow{2}{*}{w/o HP} & 36.68 & 28.96 & 30.94 \\
    ~ & \red{(-6.90)} & \red{(-5.59)} & \red{(-7.20)} \\
    \multirow{2}{*}{with HP} & 36.64 & \textbf{26.83} & \textbf{30.39} \\
    ~ & \red{(-6.94)} & \textbf{\red{(-7.72)}} & \textbf{\red{(-7.75)}} \\
    \bottomrule
  \end{tabular}
  }
\end{table}

The system is able to calculate the average token log probability (ALP) for each sample during decoding, and ALP values are usually statistically correlated with the WER of the sample, as shown in Figure \ref{fig:sample}. However, when combined with GPT during decoding, there are some ``outliers" in the left half, which corresponds to higher ALP values, indicating that these samples have a significantly higher WER. Upon examination, we found that these samples were trapped in hallucination, where a portion of the content was repeated incorrectly multiple times. However, after applying Hallucination Penalty (HP), the phenomenon of ``outliers" is significantly mitigated. Table \ref{tab:topresult} provides a summary of the WER on the high ALP test subset, where high ALP samples are given higher priority in speech pseudo-label training. It can be seen that compared to not using GPT, selecting the high ALP subset results in a significant additional decrease in WER compared to selecting all samples, and HP plays a significant role in this improvement.

\subsection{Leveraging speech data}
\label{section:multimedia}
\begin{figure}[htp]
  \centering
  \includegraphics[width=\linewidth]{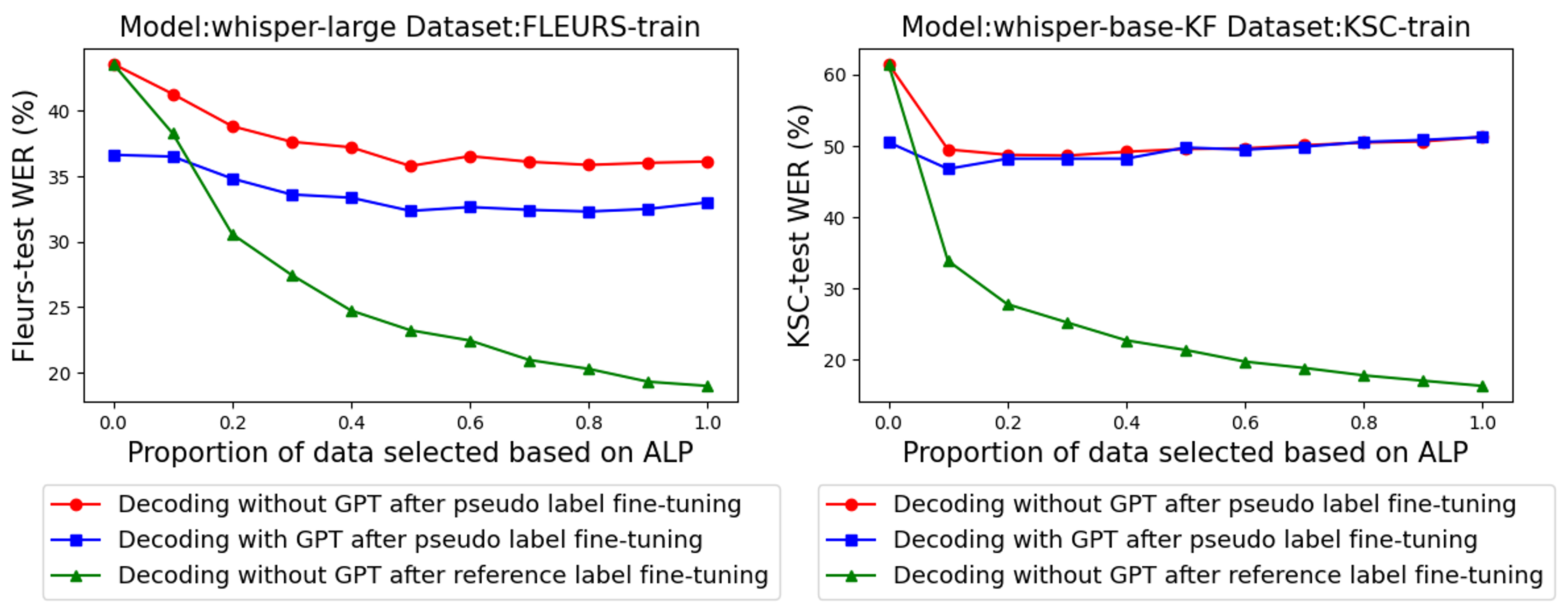}
  \caption{Relationship between the proportion of data selected based on the average token log probability and the WER of the corresponding domain test set.}
  \label{fig:relation}
\end{figure}

We used two data scales of unlabeled speech data for two models, which were first decoded with GPT to generate pseudo-labels and then used for model fine-tuning. Figure \ref{fig:relation} depicts the relationship between the proportion of data selected based on average token log probability and the corresponding WER of the corresponding domain test set. The left graph shows the results of fine-tuning on whisper-large with approximately 10 hours of FLEURS-train pseudo-labels. The right graph illustrates the results of fine-tuning on Whisper-base-KF with approximately 500 hours of KSC-train pseudo-labels. The results are presented with and without incorporating GPT decoding. Additionally, we conducted fine-tuning with manually annotated labels under the same configuration for comparative experiments to explore the difference between pseudo-labels and manually annotated labels.

From Figure \ref{fig:relation}, we observe that when performing fine-tuning with a smaller amount of data using pseudo-labels, there is a certain gain in WER reduction. However, as the selection proportion exceeds 0.5, fluctuations occur, and the WER no longer decreases significantly. Nonetheless, incorporating GPT decoding still provides some improvement. On the other hand, when fine-tuning with a larger amount of high-frequency data using pseudo-labels, there is a clear decreasing-then-increasing trend in WER after fine-tuning with varying proportions of data. This trend indicates the effectiveness of using ALP as a criterion for data selection. Furthermore, due to a sufficiently large amount of data, this pseudo-label fine-tuning method enables the model to learn language model information, leading to almost no additional improvement when decoding with GPT. This approach allows for the integration of language model knowledge into Whisper, eliminating the need for external language models during decoding and accelerating the decoding speed.

\begin{table}[h]
  \caption{Summary of the overall WER for systems leveraging unpaired speech and text data.}
  \label{tab:finalsummary}
  \centering
  \scalebox{0.82}{
  \begin{tabular}{lcc}
    \toprule
     \multirow{2}{*}{\textbf{Systems}}  & \textbf{Whisper-large} & \textbf{Whisper-base-KF}             \\
     ~ & \textbf{Fleurs-test WER} &\textbf{KSC-test WER} \\
    \midrule
    1.Origin (baseline) & 43.58\% & 61.51\% \\ 
    2.(1)+GPT for decoding & 36.64\% & 50.53\% \\
    3.Pseudo-label fine-tuning & 35.79\% & 48.66\% \\
    4.(3)+GPT for decoding & \textbf{32.36\%} & \textbf{48.23\%} \\
    \midrule
    5.Reference label fine-tuning & 23.24\% & 25.26\% \\
    Gap-filling Ratio & 0.552 & 0.366 \\ 
    \bottomrule
  \end{tabular}
  }
\end{table}

Table \ref{tab:finalsummary} summarizes the WER of systems leveraging unpaired speech and text data. The selection of system (3) corresponds to the best result among all data selection proportions. Reference label fine-tuning is performed using the same amount of data as in system (3). The Gap-filling Ratio is the ratio of the reduction in difference between system (4) and system (5) compared to system (1), which is calculated as (WER.(1)-WER.(4))/(WER.(1)-WER.(5)). In Whisper-large, by leveraging text data and unlabeled Fleurs-train speech data, we achieved an absolute WER reduction of 11.24\% on the in-domain test set. This method can achieve more than half the efficacy of reference labels, without incurring the associated human labor costs. Similarly, in Whisper-base-KF, by utilizing text data and unlabeled KSC-train speech data, we observed an absolute WER reduction of 13.28\% on the in-domain test set. Even at higher scales of data, more than one-third of the performance of the reference label can be achieved using this pipeline, resulting in a significant reduction in WER for Whisper on Kazakh.

\section{Conclusion}

In this paper, we explore how to leverage low-cost unpaired speech and text data to improve the performance of the multilingual speech recognition model Whisper on the under-represented language Kazakh. By integrating the language model GPT into Whisper's decoding process and implementing EOT judgment modification and hallucination penalty, we significantly reduce WER, particularly for samples with higher decoding average token log probability. Furthermore, we utilize this criterion to select samples for model pseudo-label fine-tuning, further improving performance. The whole process is foundational but effective in bringing low-resource languages into the wave of large speech models, and it is entirely possible to generalize it to other under-represented languages, with the potential to combine it with more novel techniques.


\bibliographystyle{IEEEtran}
\bibliography{mybib}

\end{document}